\documentclass[prl,twocolumn,showpacs]{revtex4}%
\usepackage[dvips]{graphicx}
\usepackage{epsfig}
\usepackage{psfrag}
\usepackage{amsmath}
\usepackage{amsfonts}
\usepackage{amssymb}
\usepackage{bm}%
\begin{document}
\title{Effect of nuclear quadrupole interactions on the dynamics of two-level systems
in glasses}
\author{A. L. Burin$^{\dag\spadesuit}$, I.Ya.Polishchuk$^{\spadesuit\ddagger}$, P.
Fulde$^{\spadesuit}$, Y. Sereda$^{\dag}$}
\affiliation{$^{\dag}$Department of Chemistry, Tulane University, New Orleans, LA 70118}
\affiliation{$^{\ddag}$RRC Kurchatov Institute, Kurchatov Sq. 1, 123182 Moscow, Russia}
\affiliation{$^{\spadesuit}$Max-Planck-Institut f\"ur Physik Komplexer Systeme, D-01187
Dresden, Germany}
\date{\today}

\begin{abstract}
The standard tunneling model describes quite satisfactorily the thermal
properties of amorphous solids at temperatures $T<1K$ in terms of an ensemble
of  two-level systems possessing logarithmically uniform distribution over
their tunneling amplitudes and uniform distribution over their asymmetry
energies. In particular, this distribution explains the observable logarithmic
temperature dependence of the dielectric constant. Yet, experiments have shown
that at ultralow temperatures $T<5mK$ such a temperature behavior breaks down
and the dielectric constant becomes temperature independent (plateau effect).
In this letter we suggest an explanation of this behavior exploiting the
effect of the nuclear quadrupole interaction on tunneling. We show that below
a temperature corresponding to the characteristic energy of the nuclear
quadrupole interaction the effective tunneling amplitude is reduced by a small
overlap factor of the nuclear quadrupole ground states in the left and right
potential wells of the tunneling system. It is just this reduction that
explains the plateau effect . We predict that the application of a
sufficiently large magnetic field $B>10T$ should restore the logarithmic
dependence because of the suppression of the nuclear quadrupole interaction.

\end{abstract}

\pacs{6143.Fs, 77.22.Ch,75.50Lk}
\maketitle

Low temperature properties of amorphous solids have been associated with
two-level systems (TLS) \cite{ahv, p}. Each TLS can be described by the
tunneling amplitude $\Delta_{0}$ and the two-well asymmetry energy $\Delta$.
The tunneling amplitude $\Delta_{0}$ couples two energy minima and it is
exponentially sensitive to external and internal parameters of the system.
This results in the logarithmically uniform
distribution\cite{ahv,p,Hunklinger,Phillips_review}\vspace{-0.2cm}
\begin{equation}
P(\Delta,\Delta_{0})=\frac{P}{\Delta_{0}},P=const.\label{Eq:TLSmain}%
\end{equation}
Equation (\ref{Eq:TLSmain}) leads to a universal temperature and time
dependence of various physical properties of amorphous solids, thus making
"glassy\textquotedblright\ behavior easily recognizable. One of the signatures
of the glassy state is a logarithmic temperature dependence of the dielectric
constant and the sound velocity \cite{Hunklinger,Phillips_review}.

\begin{figure}[ptb]
\centering
\includegraphics[width=3in,clip]{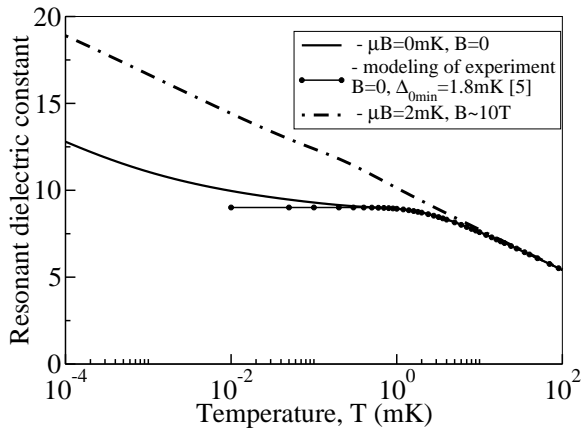}\caption{Temperature and field dependence of TLS dielectric constant. The solid line shows the temperature dependence of the dielectric constant, calculated using parameters $\lambda_{*}=4mK$, $\eta=0.33$, $n=4$ and the tunneling amplitude $\Delta_{0}$ replaced with $\Delta_{0*}=\Delta_{0}\eta^{n(1-(\Delta_{0}/\lambda_{*})^{2})^{3/2})}$ at $\Delta_{0}<\lambda_{\ast}$ as derived in Ref. \cite{large_paper}. The artificially generated solid line with circles represents experimental data \cite{DDO2}, which can be described within the standard tunneling model with the tunneling amplitude cutoff $\Delta_{0min}=1.8mK$ (representative value, see \cite{DDO2}). It is clear that our approach does not deviate from the experiment in the experimental temperature range $T>1mK$. Dielectric constant in the external magnetic field (dot-dashed line) is found using the numerical modeling for axially symmetric nuclear quadrupole interaction \cite{large_paper}.}
\label{fig:eps_h_S2}
\end{figure}

Recent experimental investigations on many amorphous solids at very low temperature
have revealed a number of qualitative deviations from the standard tunneling
model. In particular, it was demonstrated \cite{SEH,DDO2,b6,EnssReview} that
for $T\leq5mK$ the dielectric constant is approximately temperature
independent rather than logarithmically temperature dependent. These results
conflict with the logarithmically uniform distribution Eq. (\ref{Eq:TLSmain})
of TLS over tunneling amplitudes. To resolve the problem one can assume that
the distribution of TLS has a low energy cut-off at $\Delta_{0,min}\sim 1-3 mK$ (See \cite{DDO2}, Fig. \ref{fig:eps_h_S2}).
Although this assumption fits the experimental data quite good, it is in conflict with the observation of very long
relaxation times in glasses \cite{DDO1}. These times (a week or longer)
require much smaller tunneling amplitudes than $1mK$ (remember that the TLS
relaxation time is inversely proportional to its squared tunneling amplitude).

We suggest an explanation for this controversial behavior using recent work
\cite{WFE, bodea2004, iyp2005}, where it was proposed that the nuclear
quadrupole interaction affects the properties of TLS's at low temperatures due
to a mismatch of the nuclear quadrupole states in the two potential wells \cite{comm2}. The
case of higher temperatures $T>5mK$ has been considered in \cite{bodea2004,
iyp2005}. The significance of the nuclear quadrupole interaction has been
proven experimentally in glycerol glass \cite{D1}. This interaction also helps
to understand the unusual magnetic field dependence of dielectric properties
in non-magnetic dielectric glasses \cite{EnssReview,SEH}.

The nuclear quadrupole interaction splits the levels of a well and the
effective tunneling between pairs of them is $\Delta_{0\ast}<\Delta_{0},$
i.e., gets smaller than the initial value $\Delta_{0}$. We demonstrate that
this reduction is of decisive importance for TLS with tunneling amplitude
$\Delta_{0}$ smaller than the characteristic nuclear quadrupole interaction
energy $\lambda_{\ast}$. The contribution of these TLS's to the dielectric
constant is substantially suppressed. This can explain the experimental observations.

First, we review the key data concerning the contribution of tunneling systems
to the resonant part of the dielectric susceptibility for $T<50mK$
\cite{Hunklinger,Phillips_review}. This contribution can be described as the
adiabatic response of the quantum system in the ground state to a slow
external field. The application of an external electric field to the tunneling
system in the ground state produces a polarization along the field. For
instance, in the two-level model the external field $\mathbf{E}$ affects the
TLS Hamiltonian through the change in the asymmetry energy, i.e., $-\Delta
_{0}s^{x}-(\Delta-{\bm\mu}\mathbf{E})s^{z}$ \cite{Hunklinger,Phillips_review}.
The ground state has the energy $E_{g}=-(1/2)\sqrt{\Delta^{2}+\Delta_{0}^{2}%
},$ which results in the susceptibility $\chi_{g}=-\frac{\mu^{2}}{3}d^{2}%
E_{g}/d\Delta^{2}=\frac{\mu^{2}}{3}\frac{\Delta_{0}^{2}}{2(\Delta_{0}%
^{2}+\Delta^{2})^{3/2}}$.

A contribution of excited states to the susceptibility can be negative. In a
two-level system the contribution of the excited state with the energy
$E_{e}=(1/2)\sqrt{\Delta^{2}+\Delta_{0}^{2}}$ is
$-\chi_{g}.$ When the temperature is less than the minimum energy splitting
between the ground and excited state $E_{min}(\Delta_{0})=\Delta_{0},$ the
excited state contribution can be neglected. Then, the contribution of the
TLS's with given $\Delta_{0}>T$ and different $\Delta$ to the susceptibility
is
\begin{equation}
-P\frac{\mu^{2}}{3}\int_{-W}^{W}d\Delta \frac{d^{2}E_{g}}{d\Delta^{2}}=-P\frac{\mu
^{2}}{3}\left.  \frac{dE_{g}\left(  \Delta_{0},\Delta\right)  }{d\Delta
}\right\vert _{-W}^{W},
\label{eq:bur1}
\end{equation}
where $W\gg\Delta_{0},T$ is the maximum TLS asymmetry. For large enough
$\Delta~$one has $E_{g}(\Delta)\sim-\mid\Delta\mid/2$ and the integral
(\ref{eq:bur1}) equals \textbf{unity}. This result is valid for the ground
state of the tunneling system irrespective of the presence of the nuclear
quadrupole interaction. Then, one should make averaging of
$P\frac{\mu^{2}}{3}$ over $\Delta_{0}$ using the distribution Eq.
(\ref{Eq:TLSmain}) with the constraint that the minimum energy splitting
$E_{min}$ between the ground- and first excited state exceeds $T$. This is
realized when $\Delta_{0}>T$. Otherwise, the positive dielectric
susceptibility of the ground state is compensated by the negative contribution
of the excited state and the total contribution can be neglected. Then one may
express the TLS resonant susceptibility to logarithmic accuracy in the form
\begin{equation}
\chi_{TLS}\approx\frac{P\mu^{2}}{3}\int_{0}^{W}\frac{d\Delta_{0}}{\Delta_{0}%
}\Theta(E_{min}(\Delta_{0})-T),
\label{eq:log_accr}
\end{equation}
where $\Theta(x)$ is the Heaviside step function. In the standard TLS model
$E_{min}(\Delta_{0})=\Delta_{0}$ \cite{Hunklinger,Phillips_review} and we
obtain the well-known result $\chi_{TLS}\approx(P\mu^{2}/3)ln(W/T)$. Equation
(\ref{eq:log_accr}) is still applicable in the low temperature limit, where
the nuclear quadrupole interaction is significant.

To describe the effect of the nuclear quadrupole interaction on the
dielectric constant, we consider the change in TLS ground state properties
induced by this interaction. The adiabatic (resonant) dielectric
susceptibility $\chi\propto \Delta^{-3}$ (see Refs. \cite{Hunklinger,Phillips_review}) 
is due to the TLS's with small asymmetry energies $\Delta
\leq\Delta_{0}$. Moreover, as discussed previously, the minimum energy
splitting between the ground and first excited state takes place at $\Delta
=0$. Therefore, in what follows we study this particular case.

Consider a two-level system formed by $n$ tunneling atoms possessing a nuclear
spin $I\geq1$ and consequently a nuclear quadrupole moment. The tunneling
system Hamiltonian can be expressed through the standard TLS pseudospin
Hamiltonian $-\Delta_{0}s^{x}$ and the quadrupole interactions, which are
$\widehat{H}_{r}$ and $\widehat{H}_{l}$ in the left well ($s^{z}=-1/2$) and
the right well ($s^{z}=1/2$), respectively. The mismatch of the nuclear
quadrupole interactions between the two wells makes the TLS properties
sensitive to an external magnetic field \cite{WFE}. We also assume that
the choice of $\Delta=0$ implies identical ground state energies of nuclear
spins in the right and left wells, i.e., $<gl\mid\widehat{H}_{l}\mid
gl>=<gr\mid\widehat{H}_{r}\mid gr>=E_{quadr,}^{loc}$, where $|gl>$ ($|gr>$) is
the ground nuclear spin state in the left (right) well. The modified TLS
Hamiltonian takes the form
\begin{equation}
\widehat{h}_{TLS}=-\Delta_{0}s^{x}+(\widehat{H}_{r}+\widehat{H}_{l}%
)/2+(\widehat{H}_{r}-\widehat{H}_{l})s^{z}.\label{eq:tls_Ham}%
\end{equation}

Two different realizations of the ground state are possible for the
Hamiltonian Eq. (\ref{eq:tls_Ham}) depending on the relative size of the TLS
tunneling amplitude $\Delta_{0}$ and the nuclear quadrupole interaction energy
$\lambda_{\ast}$. If  $\Delta_{0}>\lambda_{\ast}$, then the hybridization is
strong and the tunneling particle is equally shared between both wells and one
can set $s^{x}=1/2$, $s^{z}\approx0$ in Eq. (\ref{eq:tls_Ham}). Then, the
ground state energy is estimated as
%\begin{equation}
$E_{g}^{hyb}=-\Delta_{0}/2+E_{quadr}^{hyb}$,
%\label{eq:symm_st}
%\end{equation}
where $E_{quadr}^{hyb}$ is the ground state energy of the nuclear spin in the
\textit{hybridized} Hamiltonian $(\widehat{H}_{r}+\widehat{H}_{l})/2$. If
$\Delta_{0}<\lambda_{\ast}$, the ground state of the system corresponds to
either the \textit{localized} left well ground state or the right one
$E_{quadr}^{loc}=~<gl\mid\widehat{H}_{l}\mid gl>=<gr\mid\widehat{H}_{r}\mid
gr>$. The tunneling term in Eq. (\ref{eq:tls_Ham}) leads to a small correction
to this energy given by
\begin{equation}
\delta E=-\eta_{\ast}\Delta_{0}/2,\ \ \ \ \ \eta_{\ast}%
=<gl|gr>.\label{eq:renorm_ampl1}%
\end{equation}
The parameter $\eta_{\ast}<1$ characterizes the average overlap integral of
the nuclear spin ground states in the left and right wells. We assume that
$\eta_{\ast}\ll1.$ For this reason, correction (\ref{eq:renorm_ampl1}) is
ignored below when estimating the energy.

The energy splitting between the ground and the first excited state, formed by
the symmetric and antisymmetric superpositions of the localized ground states
in the left and right wells is given by the modified effective tunneling
amplitude
\begin{equation}
\Delta_{0\ast}=\eta_{\ast}\Delta_{0}=E_{min}(\Delta_{0}),
\label{eq:renorm_ampl}%
\end{equation}
which also defines the minimum energy splitting of the ground and first
excited states $E_{min}(\Delta_{0})$, because other excited states can be neglected (see below Eq. (\ref{eq:pert_theor1}) and the related discussion).

We have found the two possible realizations for the ground state of TLS's
affected by the nuclear quadrupole interaction, a \textit{hybridized} and a
\textit{localized} one. For the parameter $\lambda_{\ast}$ introduced above
one has 
\begin{equation}
\lambda_{\ast}=2(E_{quadr}^{hyb}-E_{quadr}^{loc}). 
\label{eq:lamd_st2}
\end{equation}
In the case
%\begin{equation}
$\Delta_{0}>\lambda_{\ast}$,
%\label{eq:lamd_st1}
%\end{equation}
i.e., the TLS is hybridized between the left and right wells and the nuclear
quadrupole interaction can be approximately ignored. In the opposite case
%\begin{equation}
$\Delta_{0}<\lambda_{\ast}$,
%\label{eq:lamd_st2}
%\end{equation}
the nuclear quadrupole interaction localizes the tunneling particle in one of
the wells leading to a strong reduction of the effective tunneling amplitude
and replacing $\Delta_{0}$ by $\Delta_{0\ast}=\eta_{\ast}\Delta_{0}\ll
\Delta_{0}$. The nuclear spin ground state in each well can be treated as
non-degenerate \cite{note_degen}. Then, one can approximately define the
ground state and the lowest excited state as superpositions of these two
lowest energy states. The ground state of the nuclear spin in each of the two
wells can be considered as non-degenerate and one may construct the ground
state and the lowest excited state as a superposition of unperturbed
ground states in the two wells. Higher excited state contributions are
neglected. This is justified by calculating their contribution to lowest order
perturbation theory. We find a correction factor $c$ to the ground state
amplitude as
\begin{equation}
c=1-\sum_{i\neq gr}\frac{\Delta_{0}^{2}\mid<gl|i>\mid^{2}}{\left(
E_{i}-E_{gl}\right)  ^{2}}.\label{eq:pert_theor1}%
\end{equation}
We determine the parameter regime for which the second term can be neglected
as follows. Assume that each TLS contains $n$ simultaneously tunneling atoms. 
The lowest excited states in each well are separated from the
ground state by some characteristic energy $b\sim\hbar\omega_{0}$ where
$\omega_{0}$ is a frequency of the nuclear quadrupole resonance. The next
group of states is separated by the energy interval $\sim 2b$. The vast
majority of the states have energies exceeding that for the ground state by
the energy $nb\sim \lambda_{\ast}/2$ (see Eq. (\ref{eq:lamd_st2}) and \cite{large_paper}). Because of the large statistical weight, these
states provide the main contribution to Eq. (\ref{eq:pert_theor1}). So, we may
replace the denominator in Eq. (\ref{eq:pert_theor1}) by $\lambda_{\ast}$. Then, the
sum of the $\mid<gl|i>\mid^{2}$in the numerator can be approximated by
$1$. The requirement that the second term in Eq. (\ref{eq:pert_theor1}) is
small results in the condition
\begin{equation}
\Delta_{0}<\lambda_{\ast}.\label{eq:pert_theor2}%
\end{equation}
Thus $c\approx 1$ when $\Delta_{0}<\lambda_{*}$ so in that case our consideration can be restricted to the two lowest energy levels.

Consider the adiabatic dielectric
susceptibility at different temperatures. For high temperature $T>\lambda
_{\ast}$ the quadrupole interaction is not significant and we may use the
result of the standard tunneling model
\begin{equation}
\chi_{TLS}(T)\approx(P\mu^{2}/3)ln(W/T),\ \ \ \lambda_{\ast}%
<T<W.\label{eq:highTans}%
\end{equation}
Now consider the limit of low temperatures, where the main contribution to the temperature dependence comes from the TLS's with modified tunneling splitting
Eq. (\ref{eq:renorm_ampl}). According to Eq. (\ref{eq:log_accr}), the TLS
contribution to the dielectric constant can be estimated by the logarithmic
integral with the lower limit $\Delta_{0,min}$, defined by the condition
$E_{min}(\Delta_{0,min})=T$. At low temperature this minimum energy splitting
is given by the renormalized tunneling amplitude $\Delta_{0\ast}=\eta_{\ast
}\Delta_{0}$ (see Eq. (\ref{eq:renorm_ampl})) and the integral in Eq.
(\ref{eq:log_accr}) should be cut-off at $\Delta_{0,min}=T/\eta_{\ast}$ rather than at $\Delta_{0,min}=T$ as in the standard tunneling model. This leads to a negative correction to Eq. (\ref{eq:highTans}) of the form
\begin{equation}
\chi_{TLS}(T)\approx(P\mu^{2}/3)(ln(W/T)-ln(1/\eta_{\ast}).
\label{eq:lowTans}
\end{equation}
The condition for the applicability of Eq. (\ref{eq:lowTans}) is
given by the criterion Eq.(\ref{eq:lamd_st2}) which has to be valid for
tunneling system with the smallest involved tunneling amplitudes, in
particular $\Delta_{0,min}<\lambda_{\ast}$. Using $\Delta_{0,min}<T/\eta
_{\ast}$ we obtain the criterion for the validity of Eq.
(\ref{eq:lowTans})
\begin{equation}
T<\eta_{\ast}\lambda_{\ast}.
\label{eq:lowTdef}
\end{equation}
The value of Eq. (\ref{eq:lowTans}) at the highest admissible temperature
$\eta_{\ast}\lambda_{\ast}$ coincides with the value of Eq.
(\ref{eq:highTans}) at the lowest admissible temperature $\lambda_{\ast}$.
Since the adiabatic dielectric constant cannot decrease with decreasing
temperature, because excited states have a negative effect on it, 
we expects that it does not change at intermediate
temperatures (plateau regime)
\begin{equation}
\chi_{TLS}(T)\approx(P\mu^{2}/3)ln(W/\lambda_{\ast}),\ \ \ \lambda_{\ast}
\eta_{\ast}<T<\lambda_{\ast}.
\label{eq:intermTans}
\end{equation}
The plateau can also be interpreted as the consequence of the pseudogap in the 
distribution of effective tunneling amplitudes $\Delta_{0*}$ within the 
range $\lambda_{\ast}\eta_{\ast}<\Delta_{0\ast}<\lambda_{\ast}$
\cite{large_paper}.

To compare the theory with the experiment let us estimate the parameter $\lambda_{\ast}$. We can use
the approximate expression obtained above, i.e., $\lambda_{\ast}\simeq nb$
where $n$ is the number of atoms per two level system and the energy $b$ has
been defined before. The majority of glasses showing the plateau in the
temperature dependent dielectric constant
\cite{SEH,DDO2,b6} contains $K$, $Na$, $Al$, $Ba$  atoms which possess a
relatively large nuclear quadrupole moment (see \cite{comm2,large_paper}). The nuclear quadrupole splitting energy $b$ measured for these
nuclei in different environments varies between $0.4mK$ and $1mK$ \cite{large_paper}. To obtain
$\lambda_{\ast}\sim 2nb \sim 5mK$ (see Eq. (\ref{eq:lamd_st2})) one can suppose that $n\approx 4$ atoms are
involved in the tunneling motion. This estimate looks quite reasonable taking
into account the strong effect of each tunneling atom on 
its neighboring atoms. Tunneling of the single atom shifts equilibrium positions of its neighbors. During the tunneling event the neighbors move to new equilibrium positions simultaneously with the tunneling atom. Therefore they participate in tunneling and should be included into the definition of the parameter $n$, which can thus become larger than $1$.  
The overlap parameter $\eta_{\ast}=<lg|rg>$ is taken between the ground states of $n$ non-interacting nuclear
spins and estimated as
\begin{equation}
\eta_{\ast}=\eta^{n}.\label{eq:overl_1}
\end{equation}
where $\eta$ is the ground state overlap integral for one spin.

For a crude estimate of $\eta$ we consider the orientational glass
$(KBr)_{1-x}(KCN)_{x}$ \cite{KBr} where the $CN$ group rotates between
different equilibrium positions by an angle $\phi=cos^{-1}(1/3)\approx0.123$
\cite{Hunklinger}. For the sake of simplicity consider the simplified
model of a nuclear spin $I=1$ with a quadrupole interaction described by the
axially symmetric Hamiltonian \cite{WFE} $b(I_{\alpha}^{2}-2/3)$ and $b>0$.
The index $\alpha$ denotes the nuclear quadrupole axis. Then one may estimate
a single nuclear overlap integral as $\eta\approx cos(\phi)\approx0.33$. If a
TLS consists of $n=4$ tunneling atoms, the effective overlap
integral for two ground states can be estimated as $\eta_{\ast}\approx
10^{-2}.$ In this case $\eta_{\ast}\lambda_{\ast}\approx50\mu K$ and
the plateau should show up in the temperature range from $T_{down}\sim 50\mu K$ to $T_{up}\sim 4mK$ (see
Eq. (\ref{eq:intermTans})). For $T<50\mu K$ the logarithmic temperature
dependence of the dielectric constant should be restored and resemble the one
at high temperatures (see Fig. \ref{fig:eps_h_S2}). Note that the measurements of $T_{down}$ and $T_{up})$ 
will permit us to estimate the number of atoms participating in TLS as $n\sim ln(T_{up}/T_{down}%
)$. This is important for understanding the microscopic nature of
tunneling systems.

The theory can be verified by  the measurements of the dielectric constant
at low temperatures $T<\lambda_{\ast}$ under the strong magnetic field, 
affecting the nuclear magnetic moments. When the field strength becomes comparable to
the nuclear quadrupole interaction, it changes the spin quantization axes in
both left and right wells aligning the nuclear spins along the field.
Accordingly the mismatch of spin quantization axes is reduced by the magnetic
field and the overlap integral $\eta_{\ast}$ increases and approaches unity in
the large field limit. Thus we predict that a very strong magnetic field 
eliminates the influence of the nuclear quadrupole interaction on the dielectric constant and restore
the standard tunneling model behavior. The critical
magnetic field strength at which the nuclear quadrupole effects are eliminated
can be estimated by requiring that the Zeeman splitting in the field is
comparable with the energy $\lambda_{\ast}\sim 4mK$. This expectation agrees
with our numerical estimate \cite{large_paper}. Assuming that for the typical nuclear magnetic moment $\mu$ one has $\mu B \sim 1mK$ at $B\sim 5T$, one can
estimate the magnetic field necessary to eliminate the plateau in the
dielectric constant to be $B\simeq 10T$. The calculations of the temperature
dependence of the dielectric constant at zero and large $B$
\cite{large_paper} are shown in Fig. \ref{fig:eps_h_S2} together with the plot, 
representing the experimental data \cite{DDO2}.

In summary, we explained the plateau
in the temperature dependence of the dielectric constant by
the localization of the tunnelling particle by
nuclear quadrupole interactions. We predict that an external magnetic field $B>10\ T$ will
restore the logarithmic temperature dependence of the dielectric constant in
the whole temperature range. Such an experiment will verify the theory.

AB and YS are supported by the Louisiana Board of Regents
(Contract LEQSF (2005-08)-RD-A-29). IYP is supported by the program of Russian Scientific
school.

%\end{thebibliography}


\begin{thebibliography}{99}
\bibitem{comm2}
To our knowledge the only one 
glass shows no plateau in the dielectric constant, which is mylar, containing no nuclei with spin $I>1/2$ \cite{DDO2}. This experimental result and similar results in other systems support our theory \cite{large_paper}.
\end{thebibliography}
\end{document}